\documentclass[12pt]{iopart}

\usepackage[english] {babel}
\usepackage{verbatim}
\usepackage[latin1]{inputenc}
\usepackage{multirow}
\usepackage{hyperref}
\usepackage{cite}
\usepackage{mathtools}
\usepackage{amssymb}
\usepackage{graphicx}
\graphicspath{{Illustrations/}}

\begin{document}
	\title[Uncertainty evaluation of dimensional nanoparticle diameters]{Uncertainty evalutation through data modelling for dimensional nanoscale measurements}
	\author{J. P\'etry, B. De Boeck, N. Seba\"ihi, M. Coenegrachts, T. Caebergs, M. Dobre}
	
	\address{National Standards, FPS Economy, 16 Boulevard du Roi Albert II, B-1000 Brussels, Belgium}
	\ead{jasmine.petry@economie.fgov.be}
	\vspace{10pt}

\begin{abstract}
	A major bottleneck in nanoparticle measurements is the lack of comparability. Comparability of measurement results is obtained by metrological traceability, which is obtained by calibration. In the present work the calibration of dimensional nanoparticle measurements is performed through the construction of a calibration curve by comparison of measured reference standards to their certified value. Subsequently, a general approach is proposed to perform a measurement uncertainty evaluation for a measured quantity when no comprehensive physical model is available, by statistically modelling appropriately selected measurement data. The experimental data is collected so that the influence of relevant parameters can be assessed by fitting a mixed model to the data. Furthermore, this model allows to generate a probability density function (PDF) for the concerned measured quantity. Applying this methodology to dimensional nanoparticle measurements leads to a PDF for a measured dimensional quantity of the nanoparticles. A PDF for the measurand, which is the certified counterpart of that measured dimensional quantity, can then be extracted by reporting a PDF for the measured dimensional quantity on the calibration curve. The PDF for the measurand grasps its total measurement uncertainty. Working in a fully Bayesian framework is natural due to the instrinsic caracter of the quantity of interest: the distribution of size rather than the size of one single particle. The developed methodology is applied to the particular case where dimensional nanoparticle measurements are performed using an atomic force microscope (AFM). The reference standards used to build a calibration curve are nano-gratings with step heights covering the application range of the calibration curve. 
\end{abstract}

\section{Introduction}

Measurements of nanoparticles can be performed with many different instruments, from microscopy to light scattering techniques, each having its own measurand. In order to compare data from a given technique or from different techniques, it is needed to evaluate realistically the measurement uncertainty. In the absence of a physical model of the measurement itself, classical approach of uncertainty calculation as described in the Guide of Uncertainty Measurement leads to uncertainty underestimation \cite{gum}. At the nanometer scale, we are often in this case. Hence, to establish realistic traceability at the nanometer scale, innovative methods need to be developed. The present paper proposes a method for the uncertainty evaluation of isolated single particle diameter and mean particle diameter measured by microscopy. Although the particular case of Atomic Force Microscopy (AFM) is taken, the method can be applied to different microcopy techniques, such as Transmission Electron Microscopy. 

\section{State-of-the-art}
There are currently two commonly accepted approaches to perform an uncertainty evaluation : the modelling approach and the empirical approach, which are covered by the ISO 5725 standard \cite{iso5727} and the Guide of Uncertainty Measurement (GUM) \cite{gum}. In the GUM, the uncertainty evaluation is performed by propagating the various sources of uncertainty through a measurement equation. While in ISO 5725, the variability of the measurand is captured in a statistical model and a classical analysis of variance is performed to evaluate the variability of the different components. In \cite{deldossi}, the two approaches are extensively compared.

Concerning the particular case of uncertainty calculation in Atomic Force Microscopy, the uncertainty of the nanoparticles diameter has been evaluated so far with the classical GUM approach although no global physical model of the measurement exists (\cite{AFMdelval}, \cite{meli2012}). In \cite{detemmerman2014}, the ISO 5725 approach has been used for TEM measurements and in \cite{braun} for light scattering techniques. 

In the following, we further develop the ISO 5725 approach. Instead of using a stepwise approach to determine the influential parameters separately, we design an extensive experiment and statistically analyze the measurement data to quantify the different sources of variability. A more general methodology of mixed models replaces the clasical ANOVA approach and the mixed model is fitted to the experimental data using Bayesian inference.

\section{Experimental framework}

In this contribution, nanoparticle diameters are measured with Atomic Force Microscope. Among the different AFM measurands that describe nanoparticles diameter, we use the nanoparticle height of isolated particles deposited on flat surface, as defined in (\cite{meli}). The nanoparticles are gold particles RM8012 with nominal diameter of 30 nm.

Step height standards are used for the instrument calibration, with step heights covering the range of interest for the nanoparticles measurements. Standards S$_{[1]}$ have a nominal step height of 18 nm, S$_{[2]}$ of 44 nm, S$_{[3]}$ of 100 nm and S$_{[4]}$ of 180 nm.   

The uncertainty is calculated under intermediate precision conditions. Intermediate precision, also called within-lab reproducibility, is the precision obtained within a single laboratory over a long period of time. It takes into account more changes than repeatability but less than reproductibility. In the present case, various operators measure on different days, at various positions on the sample, with varying AFM critical parameters: the probe, the tapping force and the scan speed. Experiments have been designed to identify the significant parameters that contribute to the uncertainty and to quantify these individual contributions.

\section{Methods}
\subsection{Traceability route and uncertainty evaluation}

The metrological traceability to the SI units is obtained by comparison with step-height standards. A multiple points calibration curve is built by comparing the calibrated step-height of a series of reference standards with their measured value. The reference standards are certified nano-gratings. In the particular case of nanoparticles measured by AFM, stepheight standards are nano-gratings and the measured quantity to be adjusted is the  height of a single particle, $h$, and the mean height of the nanoparticle sample, $\mu$. Nevertheless the methodology presented hereafter is valid for an arbitrary dimensional quantity. 


\subsection{Mixed model instead of measurement equation}

In absence of a measurement equation, the variability of the measured quantity is modelled by an equation containing the main inluencing influencing factors. This model is chosen to be linear and contains fixed and random effects, it is a linear mixed model. Factors are considered fixed when the same value can be repeated in a subsequent experiment and random when the experimenter randomly samples the values of the factor from a population. 

The different factors that could influence the measurand have been a priori included in the model. After running a designed experiment, the significant factors have been identified and quantified, both for the measured stepheight standards and the measured nanoparticles height.

\subsection{Design Of Experiment}
\begin{figure}
\begin{center}
	\includegraphics[width=6cm]{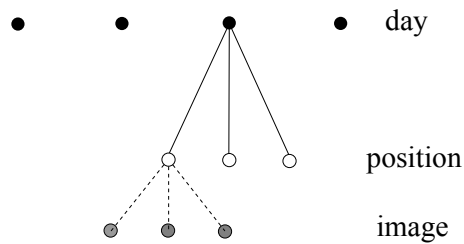}
\end{center}
\caption{Nested design schema for the random variables}
\label{nestedd}
\end{figure}

The random factors under intermediate precision conditions in microscopy are typically measurement day, measurement position and recorded image. As the levels of the different factors are not changing independently, the design is nested as depicted in figure \ref{nestedd} \cite{Montomery}. The effect of ambient conditions is minimized by working under clean and controlled laboratory conditions (stable temperature and relative humidity level, vibration damping and acoustic enclosure) and in absence of a drift effect, the deviations in ambient conditions merely result in normally distributed residual z-noise.

Considering the fixed effects, the sources of fluctuations are the operator-related settings (e.g. imaging force, scan range, scan speed, electronic feedback controller parameters, ...) and the image analysis characteristics (e.g. software used, parameter choice in used algorithms). 

Among the operator parameters, \textit{a priori} important parameters are the microscope probe, the tapping force and the scan speed. Three different types of commercial probes classically used in tapping-mode AFM have been used. The tapping force has been varied in a range corresponding to soft tapping. At last, scan speed has been varied a in range of classical use. The operator effect that may be caused by the remaining subjective choices or by any physical instrument manipulations is assessed by three operators. 

Regarding the image analysis, the software SPIP has been selected for post-processing and analysis \cite{SPIP}. This first consists in levelling the image, globally and subsequently line by line. The zero-level is fixed as the mean height of the image excluding the nanoparticules features. In a second step, in the case of nano-gratings, the ISO standard 5436-1 algorithm for step heights is applied \cite{ISO5436}. In the case of nanoparticles, the maximum $z$-value with respect to the zero-level for the isolated nanoparticles are reported. All parameter choices for these image post-processing and analysis steps are written down in a procedure, strongly decreasing the subjective choices to be made by the image analyst. Nevertheless, the possibly remaining image analyst effect is investigated as well for three analysts. In summary, the fixed effects under considerations are : the probe, the tapping force, the scan speed, the operator and the image analyst.

Table \ref{fixedeffects} summarizes the different fixed effects and the number of levels considered in the design.

\begin{table}[!ht]
	\begin{center}
		\begin{tabular}{ c | c }
			fixed effects & number of levels \\ \hline
			
			probe & 3\\ 
			tapping force & 4\\ 
			scan speed& 3\\ 
			operator&3\\
			image analyst&3\\ 
		\end{tabular}
		\caption{Fixed factors and the associated number of levels considered}
		\label{fixedeffects}
	\end{center}
\end{table}

\subsection{Probability Density Function}
Rather than the dimension of one single nanoparticle, the quantity of interest is the distribution of the nanoparticle height. Moreover, the variability intrinsic to measurement is best described by associating a random variable with the measured quantity so that its measurement result is presented by a probability density function (PDF). Such a PDF contains all available information about the measured quantity, in particular its mean, mode or median, and its standard deviation or another measure for its uncertainty. By such, the uncertainty of the measured quantity is obtained from the PDF of its associated random variable.

\subsection{Mixed model}
 
The variability of the measured quantity as a function of the main influencing fixed factors and the random effects can be described by a linear mixed model with the following equation:
\begin{equation}\label{fullmodel}
Y_{[r]ijkl}=\mu_{[r]}+\mu_{[r]i}+\mu_{[r]ij}+\mu_{[r]ijk}+\sum_{m=1}^{M}\delta_{m[r]}\mbox{X}_{m[r]ijkl}+\epsilon_{[r]ijkl},
\end{equation}
where $i=1,\ldots,I$ refers to the independent days, $j=1,..,J$ stands for the measurement positions on a given day, $k=1,\ldots,K$ refers to the repeated images for a given combination of day and position, and $l=1,\ldots,n_{ijk}$ stands for the different measurements on image $k$ taken at position $j$ on day $i$. The realization $y_{[r]ijkl}$ of $Y_{[r]ijkl}$ is thus the $l$-th observed value in the $k$-th image taken at position $j$ on the $i$-th day of the sample [r]. The random effects $\mu_{[r]i}\sim N(0,\sigma_{day}^2)$,  $\mu_{[r]ij}\sim N(0,\sigma_{pos}^2)$, $\mu_{[r]ijk}\sim N(0,\sigma_{im}^2)$ and $\epsilon_{[r]ijkl}\sim N(0,\sigma^2_{res})$ are mutually independent for all $i$, $j$, $k$ and $l$. The variance $\sigma_{day}^2$ expresses day-to-day variability, $\sigma_{pos}^2$ the variance between positions, $\sigma_{im}^2$ the variance between repeated images, and the within image variability $\sigma^2_{res}$ captures the residual variance of the observed quantity within an image. The intercept $\mu_{[r]}$ of the model represents the measured mean quantity when the fixed factors in the model have all value $0$.

The fixed effects are noted $\delta_{m[r]}$, where $m=1,..,M$ and $M$ is the number of fixed effects. Fixed effects are categorical variables and effects-type coding is used. In this case, the intercept $\mu_{[r]}$ will represent the mean response for the average value of all factors. The obtained PDF for $\mu_{[r]}$ therefore represents our knowledge about the measured mean diameter when the fixed factors are averaged out, or when there are no fixed effects present in the model (\ref{fullmodel}). Every possible level combination of the fixed factors ($\tilde{x}_m$; m=1, ...M) provides an \textit{opinion} about the measured mean diameter $\mu_{m[r]}$ ($=\mathbb{E}[Y_{[r]ijkl}]$):

$$\mathbb{E}[Y_{[r]ijkl}|\mbox{X}_{m[r]ijkl}=\tilde{x}_m, m=1,..., M]=\mu_{[r]}+\sum_{m=1}^{M}\delta_{m[r]}\tilde{x}_m$$ 

Instead of using the opinion for the average fixed factors, we make the conservative choice of combining all opinions corresponding to the different fixed factor combinations, through a linear opinion pool as in \cite{RISA187}. Assuming that all opinions are equally reliable, this means we state $\mu_{[r]}$ to be the mean of all concerning PDFs.

The estimation of the model is made with the Restricted Maximum Likelyhood method as it allows to estimate separatly the random and the fixed effects (\cite{vermol} and \cite{lind}). Fixed effects that are not significant at the $5\%$ level of significance are dropped from the model. 

\subsection{Bayesian approach}

The measured quantity is considered as a ditribution rather than a single value, hence the approach to fit the model (\ref{fullmodel}) is chosen to be Bayesian. In such framework, a fitted model can be updated by adding new data. We can repeat the above approach by performing new measurements and use the posterior PDFs for the parameters in the model (\ref{fullmodel}) as prior knowledge for the Bayesian inference of the new measurement data, taking profit from the accumulation of knowledge. The model fitting is performed by using an Hamiltonian Monte Carlo algorithm (RStan package in R software; see \cite{Stan15} and \cite{Stan17}). 

\section{Application}
In the following, the method described above is applied to the step heights in a first instance to build the calibration curve, and in a second step to the nanoparticles. Finally, the calibration is curve is applied to the distribution of the measured nanoparticle height. 

\subsection{Step height standards}
\subsubsection{Experiment under intermediate precision conditions}
$\\$
The calibration curve is constructed by fitting the measured mean step heights $\mu_{m[r]}$ to the certified mean step heights $\mu_{c[r]}$. The PDF of the measured mean step height $\mu_{m[r]}$ is obtained for each standard $S_{[r]}$ through the fitting of the measurement data obtained in a designed experiment to a linear mixed model such as described by (\ref{fullmodel}). From this fitting, the contribution of the different random and fixed effects to the variability is extracted. The estimated variance of the respective random effects, $u^2[.]$, are shown in Table \ref{res_grids_r} for the 4 standards. 

\begin{table}[!ht]
	\begin{center}
		\begin{tabular}{ c | c | c| c| c}
			effect (parameter) & $u^2[.]$ for S$_{[1]}$ &$u^2[.]$ for S$_{[2]}$ &$u^2[.]$ for S$_{[3]}$ &$u^2[.]$ for S$_{[4]}$ \\ \hline
			
			day ($\mu_{[k]i}$) &0.0075&0.0001&0.5439&1.0923\\ 
			position ($\mu_{[k]ij}$)& 0.0149&0.0016&0.1623&0.6170\\ 
			image repeat. ($\mu_{[k]ijk}$)& 0.0000&0.0003&0.0002&0.0018\\ 
			within im. var. ($\epsilon_{[k]ijkl}$)&0.0309&0.0455&0.0105&1.4797 \\ 
		\end{tabular}
		\caption{Random effects influencing the step height measurements for each grating, in $nm^2$}
		\label{res_grids_r}
	\end{center}
\end{table}

Clearly, the image repeatability does not bring any variability. For all the standards, the largest contribution comes from the within image variability, but contribution from day and position cannot be discarded.  

Regarding the fixed effects, only the probe effect is significant and strangely enough, only for gratings S$_{[1]}$ and S$_{[3]}$, but not for gratings S$_{[2]}$ and S$_{[4]}$.

\begin{table}[htb]
		\begin{center}
		\begin{tabular}{ c |c }
			& standard uncertainty ($u_{probe}$)[$nm$] \\ \hline
			S$_{[1]}$ & 0,11 \\
			S$_{[2]}$ & . \\ 
			S$_{[3]}$ & 0,25 \\
			S$_{[4]}$ & . \\
		\end{tabular}
		\caption{Standard uncertainty contribution to the step height measurements for grating S$_{[1]}$ to S$_{[4]}$}
		\label{res_grids1}
	\end{center}
\end{table}

When there is no fixed effect influence, the measured mean step height $\mu_{m[r]}$ coincides with the intercept from model (\ref{fullmodel}). Since the experiment was designed to be balanced, the standard uncertainty of $\mu_{m[r]}$, defined as $SD[\mu_{m[r]}]$, is given by:  :
$$SD[\mu_{m[r]}]=[u^2_{day}/I+u^2_{pos}/(IJ)+u^2_{im}/(IJK)+u^2_{res}/n]^{(1/2)}$$ 
where $n=\sum_{i,j,k}n_{ijk}$ is the total number of observations. 

In the case of the standards for which the probe does have an effect, the measured mean step height $\mu_{m[r]}$ is given by $\mu_{[r]} + \delta_{probe,[r]}$. The corresponding standard uncertainty is given by :
$$SD[\mu_{m[r]}]=[u^2_{day}/I+u^2_{pos}/(IJ)+u^2_{im}/(IJK)+u^2_{res}/n+u^2_{probe}]^{(1/2)}$$

 The PDF for the certified step heights $\mu_{c[r]}$ (r=1..4) are obtained assuming a normal distribution with mean and standard deviation as given in the certificate. Table \ref{res_grids} summarizes the expected values and standard uncertainty of the measured mean step height for the 4 reference gratings under consideration and the corresponding certified values.


\begin{table}[htb]\label{res_grids}
	\begin{center}
		\begin{tabular}{ c | c | c | c | c }
			grating S$_{[r]}$ & $\mathbb{E}[\mu_{m[r]}]$ [$nm$] & $SD[\mu_{m[r]}]$ [$nm$]& $\mathbb{E}[\mu_{c[r]}]$ [$nm$]& $SD[\mu_{c[r]}]$ [$nm$]\\ \hline
			$r=1$ &$15.91$&$0.13$ & $15.60$ & $0.50$\\ 
			$r=2$ & $42.15$&$0.01$& $42.30$ & $0.60$\\ 
			$r=3$ & $99.06$&$0.43$& $99.00$ & $0.60$\\ 
			$r=4$ &$177.04$ &$0.70$& $177.40$ & $0.65$ \\ 
		\end{tabular}
	\end{center}
	\caption{Expected values and standard uncertainties for the measured mean step height $\mu_{m[r]}$ and the certified mean step height $\mu_{c[r]}$ of the reference standard gratings S$_{[r]}$}
	\label{res_grids}
\end{table}

\subsubsection{Construction of the calibration curve}
$$\\$$
Having the 4 calibration points $(\mu_{m,[r]},\mu_{c,[r]})$, the calibration curve is obtained by fitting the quadratic regression equation:
\begin{equation}\label{regre}
\mu_{c[r]}=\alpha + \beta\, \mu_{m[r]} + \gamma\, \mu_{m[r]}^2 + \epsilon,
\end{equation}
The parameter $\epsilon$ is a normal deviation from the quadratic model. In the Bayesian framework, these parameters are expressed by probabilities and the PDFs for $\alpha$, $\beta$, $\gamma$ and $\epsilon$ display the variability present in the calibration points and the model uncertainty. The joint PDF for $(\alpha,\beta,\gamma,\epsilon)$ is approximated by sampling a sufficient amount of times $N$ from its distribution through the following procedure: 
\begin{itemize}
	\item take $N$ samples $(\mu_{m,[r],j},\mu_{c,[r],j})$ (for $j=1,..,N$) from the 4 calibration points
	\item calculate for $j=1,..,N$ the estimated coefficients $\alpha_j$, $\beta_j$ and $\gamma_j$, and the mean squared error $s^{2}_j$ by performing $N$ times an Ordinary Least Squares quadratic regression with regression data $(\mu_{m,[r],j},\mu_{c,[r],j})$ for $r=1,\ldots,4$
	\item simulate a random value $\epsilon_j$ from $\mathcal{N}(0,s^{2}_j)$ (for $j=1,..,N$)
	\item collect the sample $(\alpha_j,\beta_j,\gamma_j,\epsilon_j)$ 
\end{itemize}

The joint distribution of $(\alpha,\beta,\gamma,\epsilon)$ is approximated by merging the $N$ samples $(\alpha_j,\beta_j,\gamma_j,\epsilon_j)$ for $j=1,..,N$. Table \ref{cal_curve} shows the results obtained for the parameters with $N=10^6$.

\begin{table}[htb]
	\begin{center}
		\begin{tabular}{ c | c | c  }
			parameter & $\mathbb{E}$& SD \\ \hline
			$\alpha$ &$-0.2059\,nm$&$0.7724\,nm$ \\ 
			$\beta$ & $1.0025$&$0.0252$\\ 
			$\gamma$ & $0.000003\,nm^{-1}$&$0.000135\,nm^{-1}$\\ 
			$\epsilon$ &$-0.0007\,nm$ &$0.6919\,nm$\\
		\end{tabular}
	\end{center}
	\caption{Expected values and standard uncertainties for the calibration curve parameters}
	\label{cal_curve}
\end{table}

\subsection{Nanoparticles}
\subsubsection{Intermediate precision experiment}
$\\$
Similarly, an intermediate precision experiment is performed for the nanoparticles, following an appropiate design of experiment. Regarding the effects of operation, we proceed as follows. To obtain a sufficient amount of particles on a single image a scan range of $3\,\mu m\times 3\,\mu m$ is chosen. The height of a nanoparticle is measured more accurately if the pixel size is smaller. In general, a larger pixel size will lead to a larger underestimation of the real particle height. Nevertheless, practical restrictions make it infeasible to keep augmenting the number of pixels. We have therefore chosen to perform measurements with $1024\times 1024$ pixels, leading to a relatively small pixel size of about $3\,nm\times 3\,nm$. The pixel size factor is thus fixed in the experiment. The effect of scan speed in the scanning direction is assessed by considering 3 different levels (i.e. $1.8\,\mu m/s$, $3.6\,\mu m/s$ and $5.4\,\mu m/s$). Also 4 levels of amplitude ratios are tested (i.e. $65\%$, $70\%$, $75\%$ and $80\%$). The electronic feedback controller parameters are chosen to obtain a good tracking of the topography. The level of these parameters is a subjective choice going into the operator effect discussed above. 

Executing the experiments and fitting the linear mixed model (\ref{fullmodel}) as described previously, leads to the results shown in Tables \ref{res_gold_r} and \ref{res_gold_f} for the random and fixed effects. 

\begin{table}[!ht]
	\begin{center}
		\begin{tabular}{ c | c | c }
			effect & parameter & $u^2[.](nm^2)$\\ \hline
			day & $\mu_{i}$ & $u^2_{day}\coloneqq\mathbb{E}[\sigma^2_{day}]=0.16$\\ 
			position & $\mu_{ij}$ & $u^2_{pos}\coloneqq\mathbb{E}[\sigma^2_{pos}]=0.88$\\ 
			image repeatability & $\mu_{ijk}$ &$u^2_{im}\coloneqq\mathbb{E}[\sigma^2_{im}]=0.00$\\ 
			within image variability & $\epsilon_{ijkl}$ & $u^2_{res}\coloneqq\mathbb{E}[\sigma^2_{res}]=7.61$ \\ 
		\end{tabular}
	\end{center}
	\caption{Estimated variance of the random effects influencing the nanoparticle height measurements}
	\label{res_gold_r}
\end{table}

\begin{table}[!ht]
	\begin{center}
		\begin{tabular}{ c | c }
			effect &  $u^2[.](nm^2)$\\ \hline
			probe & $u_{probe} \coloneqq SD[\delta]=0.49$\\ 
			
			amplitude ratio& $u_{setp} \coloneqq SD[\eta]=0.52$\\ 
						
			scan speed & $u_{speed} \coloneqq SD[\lambda]=0.39$ \\
			operator & . \\
			analyst & . \\
		\end{tabular}
	\end{center}
	\caption{Estimated variance of the respective fixed effects significantly influencing the nanoparticle height measurements}
	\label{res_gold_f}
\end{table}


Similarly to the step height measurements, the image repeatability is good and does not bring significant variability, while the largest contribution from random effects comes from the within image variability.

The fixed factors image analyst and operator have no significant effect on the nanoparticle height measurements. Nevertheless, the levels of the fixed factors probe, amplitude ratio and scan speed do have a certain effect on the nanoparticle height measurements. The different levels of the parameters is expressed by a posterior PDFs. The average of these posterior PDFs is considered to calculate the respective standard uncertainties, as the standard deviation of these average PDFs, shown in Table \ref{res_gold_f}. The main effects of the fixed factors are given in this table, although also interaction terms (amplitude ratio$*$scan speed and probe$*$scan speed) are included in the model (\ref{fullmodel}). The respective main effects $\delta$, $\eta$ and $\lambda$ of the fixed factors probe, amplitude ratio and scan speed are obtained as it was done for step height standards. The standard deviations of $\delta$, $\eta$ and $\lambda$ are given in Table \ref{res_gold_f} by $SD[\delta]$, $SD[\eta]$ and $SD[\lambda]$, which are the main standard uncertainties caused by the respective factors. 

Using the posterior PDFs for all these parameters enables us to determine a PDF for the measured mean height $\mu_m$ of the nanoparticles sample, but also for the measured height $h_m$ of a single nanoparticle from that sample. The expected value $\mathbb{E}[\mu_{m}]$ and the corresponding standard uncertainty $SD[\mu_{m}]$ for the measured mean height $\mu_{m}$, and $\mathbb{E}[h_{m}]$ and $SD[h_{m}]$ for the measured height $h_m$ of a single particle, are given in Table \ref{res_gold}. Each level combination of the fixed factors leads to another opinion about the measured mean height $\mu_m$, but also about a future observed measured height $h_m$ of a single particle. These opinions are averaged out with equal weights by taking their mean PDF to obtain a PDF for $\mu_m$ ($\mathbb{E}[Y_{ijkl}]$), and one for $h_m$, which is used to derive the results in Table \ref{res_gold}. \\

\begin{table}[htb]
	\begin{center}
		\begin{tabular}{ c | c | c | c }
			$\mathbb{E}[\mu_{m}]$ [$nm$] & $SD[\mu_{m}]$ [$nm$]& $\mathbb{E}[h_{m}]$ [$nm$]& $SD[h_{m}]$ [$nm$]\\ \hline
			$23.40$&$1.19$ & $23.39$ & $3.18$\\ 
		\end{tabular}
	\end{center}
	\caption{Expected values and standard uncertainties for the measured mean height $\mu_{m}$ and the measured height $h_{m}$ of the gold nanoparticle sample}
	\label{res_gold}
\end{table}

\subsubsection{Calibration curve}
$\\$
The regression model (\ref{regre}) expresses the relation between the measured and the certified mean step height of the reference standards, so adopting this relation to obtain a calibration curve leads to :
\begin{equation}\label{meas_model1}
q_c=\alpha + \beta\, q_m + \gamma\, q_m^2 + \epsilon,
\end{equation}
where $q_c$ is the certified quantity's random variable and $q_m$ is the measured quantity's random variable. To extract a PDF for the measurand $q_c$ by reporting a PDF for $q_m$ on the calibration curve, we use  Monte Carlo simulation.

\subsubsection{Uncertainties}
$\\$
The PDFs for certified quantities $\mu_c$ and $h_c$ are obtained by using the calibration curve once with $q_m=\mu_m$ and $q_c=\mu_c$, and once with $q_m=h_m$ and $q_c=h_c$. The PDFs main parameters are summarized in Table \ref{cer_gold}, although the most complete measurement results are the PDFs themselves. \\

\begin{table}[htb]
	\begin{center}
		\begin{tabular}{ c | c | c | c }
			$\mathbb{E}[\mu_{c}]$ [$nm$] & $SD[\mu_{c}]$ [$nm$]& $\mathbb{E}[h_{c}]$ [$nm$]& $SD[h_{c}]$ [$nm$]\\ \hline
			$23.25$&$1.44$ & $23.24$ & $3.28$\\ 
		\end{tabular}
	\end{center}
	\caption{Expected values and standard uncertainties for the certified mean height $\mu_{c}$ and the certified height $h_{c}$ of the nanoparticle sample}
	\label{cer_gold}
\end{table}

\section{Conclusions}

Traceable measurement of nanoparticles is obtained through comparison to standards. In the range of interest, a calibration curve is built on the measured and certified data of the standards. A complete measurement uncertainty evaluation of the measured dimensional quantities of the reference standards and the nanoparticles is performed. It consists in determining a PDF for the measured dimensional quantity by statistically modelling appropriately selected experimental measurement data. 

To illustrate our methodology we have applied the entire approach on the AFM measurement procedure we execute in our laboratory to obtain PDFs for the certified height of a single particle from a sample of golden particles and the certified mean height of that sample. 

The experimental data is collected via a designed experiment performed under intermediate precision conditions. The influence of all relevant parameters is assessed by fitting a Bayesian mixed model to the data. The random effects of the model are measurement day, position and image repeatability. For this sample of nanoparticles it has been demonstrated that the fixed factors scanning speed, amplitude ratio and the type of probe all have a significant effect on the nanoparticle height measurements. The relevant random effects express day-to-day variability, variability between measurement positions, between image variability and within image variability. \\

After calibration of the measured dimensional quantity of the nanoparticles, the resulting PDF for the measurand comprises the expected value of the measurand and the associated standard uncertainty.

\section{References}
\bibliography{BibNPSize,BibUncertainty}
\bibliographystyle{plain}

\end{document}